\documentclass[aps,pre,twocolumn,nofootinbib,superscriptaddress,showpacs,showkeys]{revtex4-1}

\usepackage{amssymb,amsbsy,color}
\usepackage{graphicx}
\usepackage{amsmath}
\usepackage{bm}
\usepackage{newtxtext,newtxmath}
\usepackage{xfrac}
\usepackage{amsfonts}
\usepackage{pgfplotstable}
\usepackage{filecontents}
\usepackage{longtable}
\usepackage{gensymb}
\usepackage{hyperref}
\usepackage{csvsimple}

\newcommand{\Dod}{D_{\rm OD}}
\newcommand{\Dfix}{\langle D \rangle_{\rm fix}}
\newcommand{\Dran}{\langle D \rangle_{\rm ran}}

\begin{document}

\title{Exploring the relationship between the spatial distribution of roads and universal pattern of travel-route efficiency in urban road networks}

\author{Minjin Lee}
\affiliation{Department of Applied Physics, Hanyang University, Ansan 15588, Korea}
\author{SangHyun Cheon}
\email{scheon@gmail.com}
\affiliation{Department of Urban Planning and Design, Hongik University, Seoul 04066, Korea}
\author{Seung-Woo Son}
\email{sonswoo@hanyang.ac.kr}
\affiliation{Department of Applied Physics, Hanyang University, Ansan 15588, Korea}
\affiliation{Department of Applied Artificial Intelligence, Center for Bionano Intelligence Education and Research, Hanyang University, Ansan 15588, Korea}
\author{Mi Jin Lee}
\email{mijinlee@hanyang.ac.kr}
\affiliation{Department of Applied Physics, Hanyang University, Ansan 15588, Korea}
\author{Sungmin Lee}
\email{jrpeter@gmail.com}
\affiliation{R\&D Center, PharmCADD Co., Seoul 06180, Korea
 }%

\date{\today}

\begin{abstract}
Urban road networks are well known to have universal characteristics and scale-invariant patterns, despite the different geographical and historical environments of cities. Previous studies on universal characteristics of the urban road networks mostly have paid attention to their network properties but often ignored the spatial networked structures. To fill the research gap, we explore the underlying spatial patterns of road networks. In doing so, we inspect the travel-route efficiency in a given road network across 70 global cities which provides information on the usage pattern and functionality of the road structure. The efficiency is quantified by the detour patterns of the travel routes, estimated by the detour index (DI). The DI is a long-standing popular measure, but its spatiality has been barely considered so far. In this study, we probe the behavior of DI with respect to spatial variables by scanning the network radially from a city center. Through empirical analysis, we first discover universal properties in DI throughout most cities, which are summarized as a constant behavior of DI regardless of the radial position from a city center and clear collapse into a single curve for DIs for various radii with respect to the angular distance. Especially, the latter enables us to know the scaling factor in the length scale.  We also reveal that the core-periphery spatial structure of the roads induces the universal pattern, which is supported by an artificial road network model. Furthermore, we visualize the spatial DI pattern on the city map to figure out the city-specific characteristics. The most and least efficient connections of several representative cities show the potential for practical implications in analyzing individual cities.
\end{abstract}

\maketitle


\section{\label{sec:level1}Introduction}
Urban roads are essential infrastructure for human socioeconomic activities in a city. Accordingly, various statistical metrics have been developed in recent decades to investigate the structure and performance of urban road networks~\cite{LAMMER200689,0295-5075-91-1-18003,Crucitti:2006haa,TRAVENCOLO200889,Barthelemy20111,Lee2017,chan2011urban, CHEN2019122}. The journey to establish the proper metrics has been based on the hypothesis that the urban roads are constructed with spontaneous universal patterns across cities. Therefore, a series of previous studies have focused on capturing the universal characteristics of road networks in different cities of different sizes, and found the various scaling relations such as in the topological and geographical distance, and the betweeness centrality distribution ~\cite{PhysRevE.73.026130,LAMMER200689,chan2011urban,kirkley2018betweenness,strano2017scaling}. They have provided a good macroscopic overview of the universal structure regarding both the network structure and geometric properties. However, such studies relatively have a lack of careful consideration of the internal spatial structure of urban road networks, i.e., the structure-specific feature~\cite{delloye2020alonso}.

Meanwhile, the urban road networks are deeply tangled with the urban spatial structures. Despite the wide variety of spatial forms from the different urban contexts and histories, the backbone of the structure is mostly based on the monocentric framework ~\cite{delloye2020alonso,anas1998urban,broitman2020attraction,clark1951urban}, where urban activities are densified and clustered around a city center. Under this framework, a city radially expands its spatial boundaries from a central area, thus exhibiting a particular density gradient from the center in terms of population and socio-economic activities ~\cite{delloye2020alonso,clark1951urban,lemoy2020evidence, jiao2015urban,batty1992form,liu2021studying}. Recently, several attempts have been made to understand the universality embedded in the spatial patterns of urban socioeconomic attributes. The regarding studies ~\cite{lemoy2020evidence, delloye2020alonso} have found the scaling relationship in the internal radial profiles of population density and land use share across cities of different sizes, thus empirically proven the universal mechanism exists in urban spatial structure. Their findings have provided another perspective that bridges urban internal structure (e.g. density gradient) and macroscopic inter-city universal patterns (e.g. scaling relationship). Nonetheless, such studies are still limited to the spatial dependence of non-spatial attributes rather than the spatially networked structure of the road itself, so we aim to expand the strand of research toward the urban infrastructure domain and enrich the conventional understanding of universal behaviors of urban road networks. Using the radial approach as a lens, we explore the internal spatial structure of urban road networks and uncover the universal behavior embedded in the spatial patterns. 

In this study, we investigate the urban road networks across 70 global cities in different countries and different continents. For the analysis from the structural point of view, we focus on a travel route on a given road network and then use the detour index (DI), which is a useful measurement to better understand the usage pattern and functionality of road networks~\cite{aldous2010}. The DI of a given pair of origin and destination (OD) in a route is defined as the ratio between the shortest distance through the road network and the Euclidean distance, which indicates how far one should travel through the network compared to the Euclidean distance~\cite{Ganine1701079,aldous2010, Crucitti:2006haa,PhysRevE.71.036122,Levinson2009732,morphogenesis,JUSTEN2013146,Wang:2011ds,doi:10.1068/b32045, warnes_detour,MERCHAN202038}. Thus, the efficient travel route along with a straight-like line has a small DI. According to observations from the existing literature, it seems that the DI shows two characteristic features: First, most road networks show the decreasing pattern as the Euclidean distance increases~\cite{aldous2010,giacomin2015,Levinson2009732,JUSTEN2013146,Rietveld1999,MERCHAN202038,yang2018universal}. Second, most cities display similar spatial patterns of the DI~\cite{Crucitti:2006haa,MERCHAN202038,yue2019exploring,cubukcu2020}. However, the spatial information of travel routes, e.g., the vicinity of or the longinquity from a city center, is less taken into account for these studies when measuring the DI.  

To systematically analyze the intra-city spatial behavior of the DI, we characterize the routes with spatial variables (i.e., radial and angular distance) by introducing the radial approach and elaborate on how detour patterns change with the properties of routes. For further understanding of the mechanism behind the observed empirical results, we compare the observations with simulation results on artificial road networks. The key findings from our empirical and model analysis are the universal behavior in the spatial patterns of the DI depending on a spatial variable and its relationship to the intrinsic spatial structure of road networks, that is, the core-periphery structure. Additionally, we visualize the spatial distribution of the DI on maps showing the most efficient or least efficient connections of OD pairs in several representative cities. The visualization helps to understand the implication of our findings and presents exceptional cities that have different DI patterns.

The rest of this paper is organized as follows. We briefly review the sampling of OD, the concept of the DI, and the artificial road network model in Sec.~\ref{sec:detourindex}. Section~\ref{sec:results} elaborates the universal patterns of DI found from the empirical data and the model networks. Lastly, we summarize this paper, giving a brief discussion in Sec.~\ref{sec:discussion}.

\section{\label{sec:detourindex} Data and Methodology}

Before addressing the spatial properties of the detour index (DI), we illustrate the definition of the DI, after introducing the sampling method of the origin-destination (OD) pair in a city. Subsequently, we briefly review the statistical properties of DI from our dataset and finally propose a model of an artificial road network for understanding the empirical results.

\subsection{\label{sec:sampling}Radius-fixed travel route sampling}

\begin{figure}
\begin{center}
\includegraphics[width=\columnwidth]{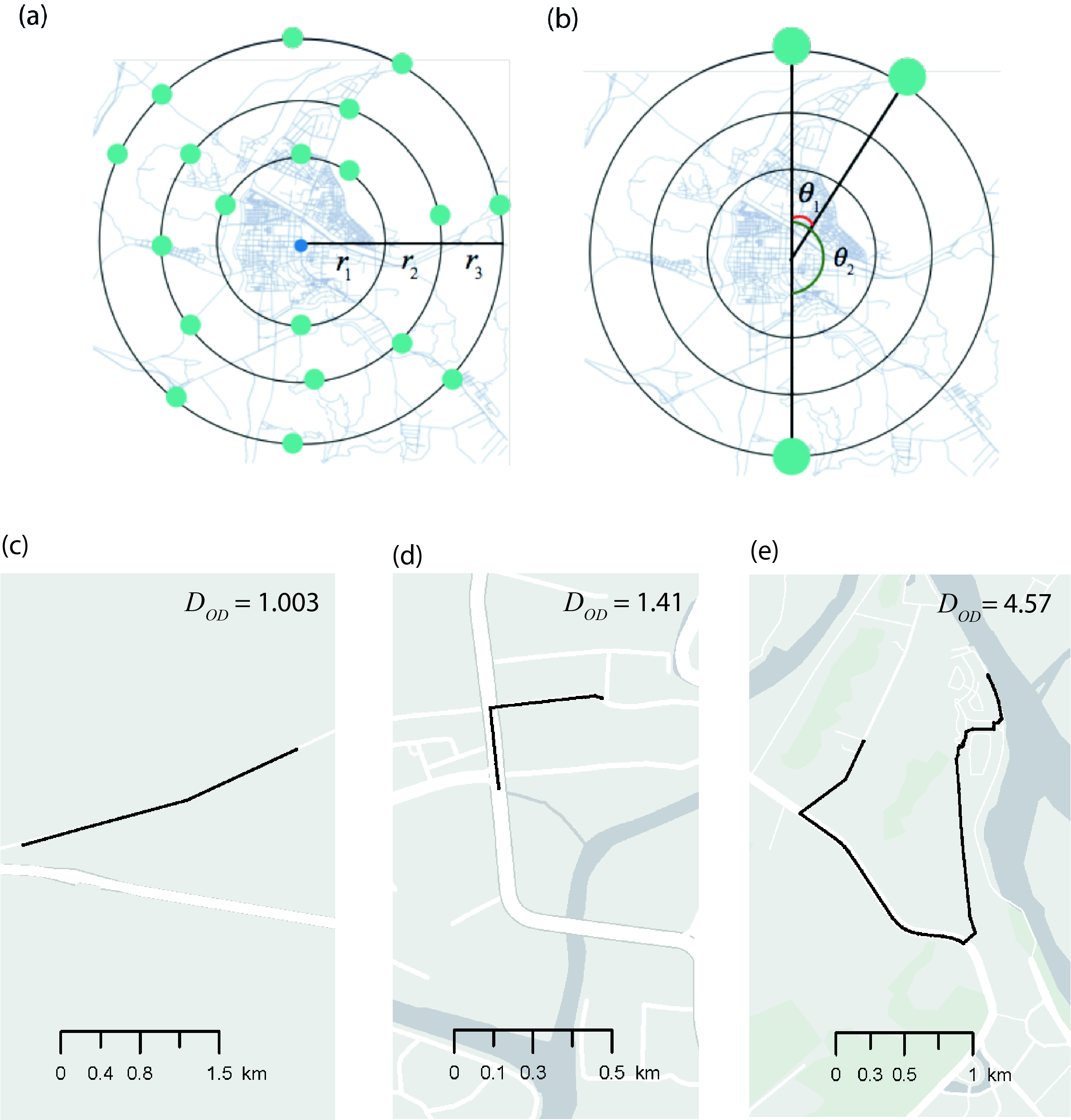}
\end{center}
\caption{Sampling of OD pairs and examples of the different detour index $\Dod$. (a) The OD points are randomly selected on a circle of a given radius $r_n$. For instance, one point on the $r_1$ and another point on the $r_2$ are not regarded as the OD pair. (b) The angular distance $\theta_{\rm OD}$ is defined as the included angle between two lines connecting the OD points from the city center. (c)--(e) Examples of travel routes with the different $\Dod$.}
\label{fig:samplingOD_DI}
\end{figure}

In our empirical analysis, we use a collection of $1\,529\,024$ travel routes in the 70 most populous global cities (the city names are listed in Table~S1 in Supplemental Material), which is composed of about $24\,000$ shortest travel routes per city. To collect the travel routes, we retrieve the shortest driving route only taken along given road networks from one place to another place (i.e., an OD pair) using {\it OpenStreetMap API}~\cite{OpenStreetMap}. We employ the radius-fixed sampling method~\cite{Lee2017} for an ensemble of travel routes to systemically apply the radial approach and analyze the DI with the spatial characteristics. In the radius-fixed sampling, the departure and arrival are chosen on one of the concentric rings from a city center as shown in Fig.~\ref{fig:samplingOD_DI}(a). Thus, for a given radius $r$, the OD points are located at the same Euclidean distance from the center. We consider the monocentric framework by giving the 70 centers for 70 cities, and the positions of the city centers are referred to the website -- {\it latlong.net}~\cite{latlong}, which provides the representative coordinates. From the six concentric rings with various radii $r_n$ ($2 \mathrm{km}$, $5 \mathrm{km}$, $10 \mathrm{km}$, $15 \mathrm{km}$, $20 \mathrm{km}$, and $30 \mathrm{km}$), we extract around $4\,000$ OD pairs on each circle, resulting in about $24\,000$ travel routes for a city.

When connecting the respective OD points to the city center straightly, the travel route can be characterized by two spatial variables --- the radius $r$ and the included angle $\theta$ called the angular distance. The angular distance is obtained from the two straight lines, as shown in Fig.~\ref{fig:samplingOD_DI}(b). Thus, $\theta \in [0, \pi]$. The small $\theta$ close to zero means the travel routes of the OD pair moving along the ring which is the city peripheries in reality, e.g., $\theta_1 \ll 1$. The large $\theta$ close to $\pi$ represents the OD pair of travel routes traversing across the city center, e.g., $\theta_2=\pi$. The Euclidean distance $s$ between the OD is $s = r\sqrt{2(1- \cos \theta)} = 2 r \sin \theta/2$ by the law of cosines. 
The radius-fixed samples are compared to those obtained by the random sampling method, which generates $20\,000$ OD pairs by connecting two random points within the city boundary of $30\mathrm{km}$ from every city center.

\subsection{\label{sec:DI_measure} Detour index (DI)}

The detour index (DI) of a given OD pair denoted by $\Dod$ is calculated as~\cite{Levinson2009732}
\begin{equation}
\Dod \equiv \frac{d_{\rm OD}}{s_{\rm OD}},
\label{eq:DI}
\end{equation}
where $d_{\rm OD}$ is the distance of the shortest travel route through the road network, and $s_{\rm OD}$ is the Euclidean distance in two dimensions. Thus, $\Dod \geq 1$ by definition. In general, $\Dod$ is not symmetric due to the existence of the one-way street between the two positions. Three examples of the DIs for different routes are displayed in Figs.~\ref{fig:samplingOD_DI}(c),~\ref{fig:samplingOD_DI}(d), and \ref{fig:samplingOD_DI}(e).
One can also obtain the various averaged value of $\Dod$ for the specific OD pairs under the constraint on $r, s,$ or $\theta$ as
\begin{align}
\mathcal{D}_r &= \frac{1}{N_{r}}\sum_{(O,D)| r_{\rm OD}=r} \Dod, \label{eq:aveDr}\\
\mathcal{D}_s &= \frac{1}{N_{s}}\sum_{\substack{(O,D)|\\ s\leq s_{\rm OD}<s+\Delta s}} \Dod, \label{eq:aveDs}\\
\mathcal{D}_{\theta} &= \frac{1}{N_{\theta}}\sum_{\substack{(O,D)|\\ \theta \leq \theta_{\rm OD} <\theta+\Delta \theta}} \Dod, \label{eq:aveDtheta}
\end{align}
where $N_r, N_s$, and $N_\theta$ represent the numbers of the radius-fixed OD pairs satisfying the respective constraints, and $r_{\rm OD}, s_{\rm OD}$, and $\theta_{\rm OD}$ are the radius from the center, the Euclidean distance, and the angular distance, respectively. The intervals of $s$ and $\theta$ are $\Delta s= 1 \mathrm{km}$ and $\Delta \theta=4.5\degree = \frac{4.5}{180}\pi \, \mathrm{rad}$.



\begin{figure*}
\begin{center}
\includegraphics[width=0.9\textwidth]{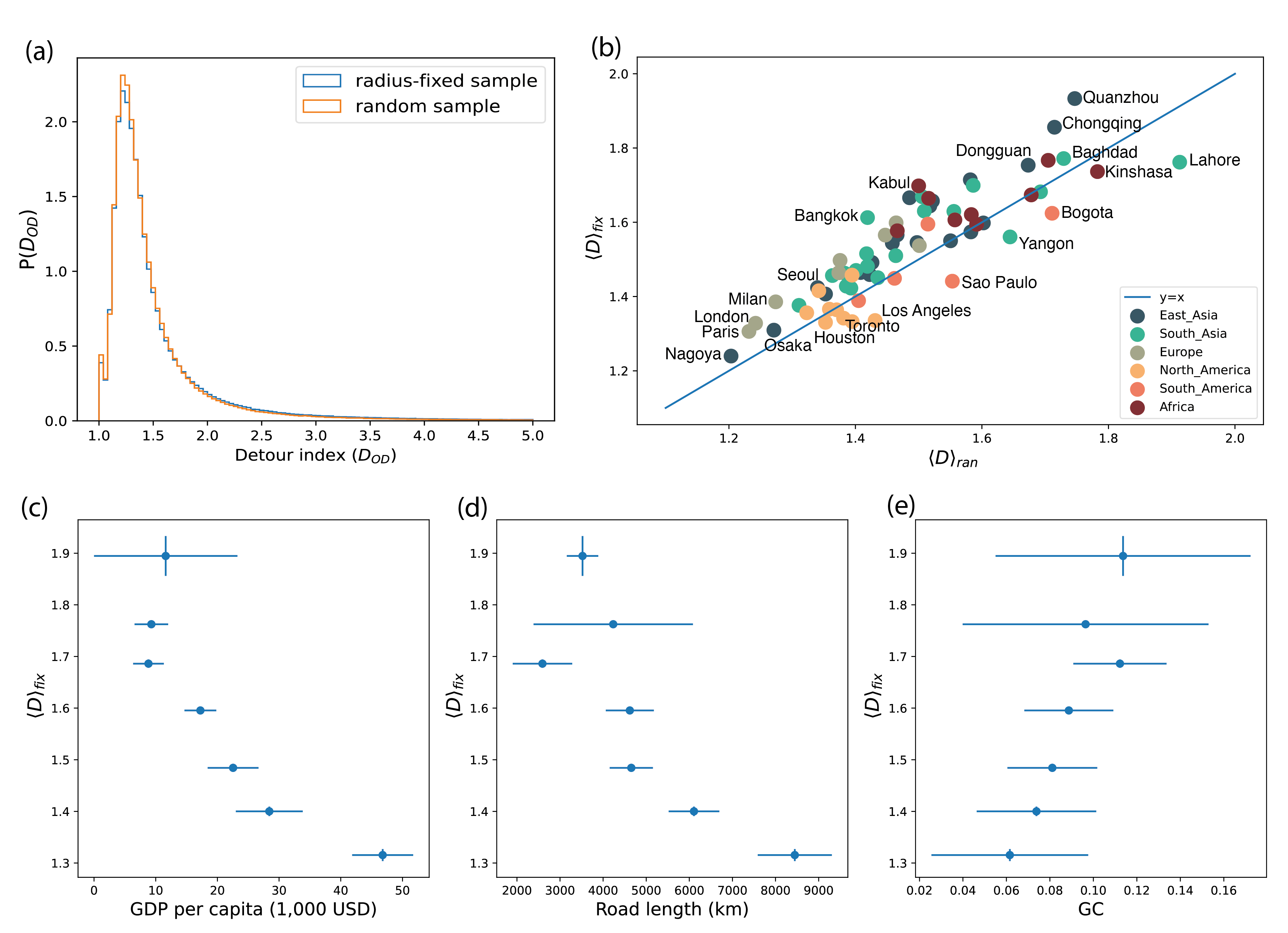}
\end{center}
\caption{Basic statistics of the detour index (DI), $\Dod$. 
(a) The distribution functions of $\Dod$ for the radius-fixed sampling and $\Dod^{\rm ran}$ for the random sampling. Each distribution is plotted from all the DI values obtained in the 70 cities, and the two distributions are almost the same. The modes in both distributions are about 1.4, which is almost similar to the averaged value $\mathcal{D}_r$ in Fig.~\ref{fig:averageDIs}(a). (b) The ensemble-averaged DI of random sampling $\langle D \rangle_{\rm ran}$ and radius-fixed sampling $\langle D \rangle_{\rm fix}$ for each city. The color code represents the different continental groups, i.e., navy for East Asia, green for South Asia, olive for Europe, yellow for North America, orange for South America, and burgundy for Africa. From the panels (a) and (b), one can infer that the distributions of $\Dod$ for individual cities are similar to that of $\Dod^{\rm ran}$ and that their shape is right-skewed unimodal mostly. The different patterns of the efficient connections of the five cities such as Seoul, London, Paris, Los Angeles, and Huston are illustrated in Figs.~\ref{fig:local_efficiency_map} and~\ref{fig:local_efficiency_map_4}, which is not captured by this macroscopic measurement. Relationship between $\langle D \rangle_{\rm fix}$ and other city indices: (c) the gross domestic product (GDP) per capita, (d) road length (we extract a road network within $30~\mathrm{km}$ radial boundary from a city center and measure the total length of road segments), and (e) geographical constraint (GC). Every point is an averaged value over the given $xy$ bins and includes the $xy$ error bars (standard error).}
\label{fig:statisticsDI}
\end{figure*}

To have a basic understanding of our DI samples, we have a look into a few statistics. First, we plot the distribution function $P(\Dod)$ in aggregate across cities [Fig.~\ref{fig:statisticsDI}(a)] which shows the right-skewed distribution with a single peak around 1.3 and 1.4. Approximately 87\% of the DIs are in the range of $1.0$ and $2.0$. The distribution says that the extremely inefficient route with $\Dod \simeq 5$ such as in Fig.~\ref{fig:samplingOD_DI}(e) is very rare and an exceptional case. Thus, we exclude the outliers and only use the routes with $\Dod < 5$ in our analysis. For both random sampling and radius-fixed sampling, there is no clear difference between the two distributions $P(\Dod)$ and $P(D_{\rm OD}^{\rm ran})$ where $D_{\rm OD}^{\rm ran}$ is the DI of an OD pair obtained from the random sampling (not considering the fixed radius). 

We also investigate two distributions of $\Dod$ and $D_{\rm OD}^{\rm ran}$ of the respective cities by comparing the mean values of a city denoted $\Dfix\equiv\sum_{r}\sum_{(O, D)|r_{\rm O}=r_{\rm D}=r}D_{\rm OD}P(D_{\rm OD})$ where $r_{\rm O (D)}$ is the radius from the center to the origin (destination), and $\Dran\equiv\sum_{(O, D)}D_{\rm OD}^{\rm ran}P(D_{\rm OD}^{\rm ran})$. Figure~\ref{fig:statisticsDI}(b) shows a scatter plot of the two mean values for 70 global cities, where the color codes represent different regional groups of East Asia (blue), South Asia (green), Europe (olive), North America (yellow), South America (orange), and Africa (brown). $\Dfix$ and $\Dran$ show almost the linear relationship and
the mean values range between $1.2$ and $2.0$ in accordance with the probable range in Fig.~\ref{fig:statisticsDI}(a). Accordingly, one can guess that every city has two distributions of $\Dod$ and $D_{\rm OD}^{\rm ran}$ almost similar to the distributions in the entire cities in Fig.~\ref{fig:statisticsDI}(a). 

The DI is related to the quality of the road infrastructure and urban environments~\cite{Lee2017}.  In Figs.~\ref{fig:statisticsDI}(c), ~\ref{fig:statisticsDI}(d), and \ref{fig:statisticsDI}(e), we further display the relationships between $\Dfix$ and other characteristics of cities: the gross domestic product (GDP) per capita as a socioeconomic index, the total lengths of the road as an infrastructure index, and the level of geographical restriction as an environmental index~\cite{Lee2017}.
A city with a high GDP per capita and well-connected road infrastructure shows lower $\Dfix$ [see Figs.~\ref{fig:statisticsDI}(c) and \ref{fig:statisticsDI}(d)], showing the negative correlations. In a network with longer total road lengths (equivalently high road density, here), it is more probable to have diverse paths including directly connected paths. Then, the routes in the dense network are likely to take more straightly connected paths, and the DI becomes lower in a natural manner. Establishing such a well-connected infrastructure is normally related to the wealth of the city. Furthermore, when there are larger geographical barriers (e.g., mountains or rivers) within the city boundary, travel routes naturally detour more because roads cannot freely pass over them.

\subsection{Road network model}
\label{sec:modelnetwork}

To find the underlying mechanism behind the pattern shown in the empirical data, we consider artificial road network simulation later in this study. We adopt and modify a popular network model of the urban street pattern in Ref.~\cite{PhysRevLett.100.138702}, so briefly review this model.

It is a network growth model and has a node and an edge representing an intersection and a road segment, respectively. We construct the starting network with ten nodes in such a way that the positions of the nodes and the links are randomly assigned in a two-dimensional $L\times L$ plane so that the network forms a connected component. The road network evolves by adding new multiple nodes at the positions drawn from a given distribution $p(\vec{r})$ where $\vec{r}$ is a position vector from a city center. We try to connect the newly introduced nodes to the existing network at each sweep consisting of the following attempts. (1) A newly introduced node tries to connect the closest point on the closest existing edge. If the closest point is not a preexisting intersection (node), the point becomes a new intersection (i.e., another new node) on the existing edge. Thus, an edge together with one more node depending on the position of the closest point is newly generated. (2) If the two newly introduced nodes share the same closest point, we take a cost-reduction action. Creating a midpoint intersection (another new node) between the two nodes after linking the two, the midpoint called a bridge node tries to be connected to the common shared point. Eventually, three edges and one more bridge node emerge newly. (3) To avoid the unrealistic tree-like structure of the road, we try to link the remaining leaf node (whose degree is 1) with the nearest leaf node, which makes each degree become two.  We control the maximum length of the edge by the length threshold $h$, so all the aforementioned attempts succeed as long as the edge to be formed has a length less than and equal to $h$. Otherwise, the attempt is rejected. Some examples of the artificial road network are shown in Figs.~\ref{fig:model_result}(a) and~\ref{fig:model_result}(b), and Fig.~S1.

The connection process involves the possibility of the additional creation of nodes and edges by chance, so it makes it hard for us to precisely control the number of nodes and edges as we wish, but is capable of doing it statistically by the distribution $p(\vec{r})$ of the position per area (density) and the length threshold $h$ for a given $L$. We set $L=60$ and the distribution as a power law 
\begin{equation}
p(\vec{r})\sim r^{-\beta},
\label{eq:position_powerlaw}
\end{equation} where $r\equiv |\vec{r}|$. This power-law distribution of intersection density characterized by the exponent $\beta$ is based on empirical observation. The average value of $\beta$ over seventy cities is $0.76(8)$ with a standard deviation of $0.23(0)$ (see Supplemental Materials for Fig.~S6), and the average $\beta$ in the model networks is about $0.93(2)$ with a standard deviation of $0.11(2)$, giving a very heterogeneous distribution similarly to empirical data. We also consider the uniform distribution for comparison, that is, $p(\vec{r})\sim 1$ (equivalently $\beta=0$).

The high $h$ allows us to connect a pair of distant nodes. The long-range connection enables the peripheries to be connected directly, representing an outer ring road in the real world. We increase  $h$ from 2 to 6 by 1. The artificial road network shows statistical properties similar to the real one, e.g., the road length or cell area distribution, and the increasing trend of the total length of the road with the increasing number of intersections, which are mentioned as intrinsic properties of the urban street network~\cite{LAMMER200689}. The basic statistics of the artificial model networks are summarized in Fig.~S2.

\section{Results}
\label{sec:results}
\subsection{\label{sec:averagedi} Scaling collapse of the average detour index into a single curve}

\begin{figure*}
\begin{center}
\includegraphics[width=1.\textwidth]{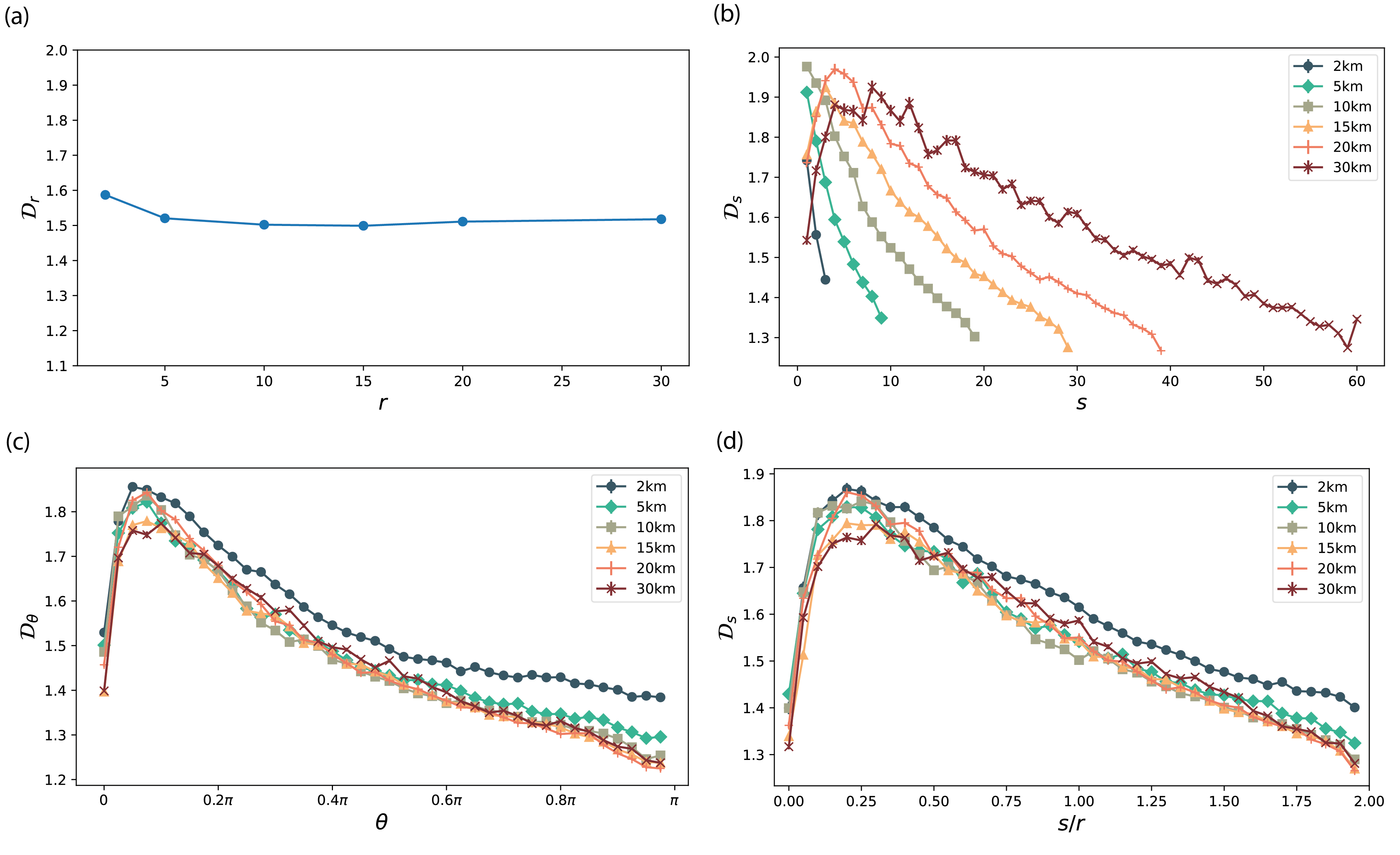}
\end{center}
\caption{Average detour index  versus the respective spatial variable over 70 cities.  (a) $\mathcal{D}_{r}$ [Eq.~(\ref{eq:aveDr})] with respect to the radius from the city center $r$, (b) $\mathcal{D}_{s}$ [Eq.~(\ref{eq:aveDs})] with respect to the Euclidean distance $s$, and (c) $\mathcal{D}_{\theta}$ [Eq.~(\ref{eq:aveDtheta})] with respect to the angular distance $\theta$, respectively. The radius-fixed sampling provides the various curves for various radii (from $2~\mathrm{km}$ to $30~\mathrm{km}$) in $\mathcal{D}_s$ and $\mathcal{D}_{\theta}$. Thanks to the nice collapse of $\mathcal{D}_{\rm \theta}$, one can find that (d) the $\mathcal{D}_{s}$ collapse to a single curve with the scaling factor $r$ easily estimated by the  trigonometric identity. The error bar represents the standard error and is about the size of the symbol.}
\label{fig:averageDIs}
\end{figure*}

As mentioned in Sec.~\ref{sec:sampling}, we analyze the spatiality of the travel-route efficiency by measuring the DI for the radius-fixed sampling with the six different radii $r=2, 5, 10, 15, 20$, and $30~\mathrm{km}$. To observe the entire pattern, we aggregate DI data across the 70 cities and investigate the averaged values from Eq.~(\ref{eq:aveDr}) to Eq.~(\ref{eq:aveDtheta}). Figure~\ref{fig:averageDIs} (a) firstly shows the pattern of $\mathcal{D}_r$, which is, in other words, the basic radial profile of DI. Interestingly, $\mathcal{D}_r$ display the almost constant behavior of around $1.5$ with respect to the radius $r$. It is completely different from the results from other urban attributes which show the decreasing patterns with the radius ~\cite{clark1951urban,lemoy2020evidence,delloye2020alonso,anas1998urban}. The constant $\mathcal{D}_r$ means that detouring behavior can only change within a given radial boundary [referring to Figs.~\ref{fig:averageDIs}(b) and~\ref{fig:averageDIs}(d)], but the overall detouring level keeps constant as the average one, independent of the radial position from the center. The example of the travel route in Fig.~\ref{fig:samplingOD_DI}(d) has a $\Dod$ close to the average value $\mathcal{D}_r$. 

Next, we explore the spatial change of the DI more specifically for a given $r$, using $\mathcal{D}_s$ as a function of the Euclidean distance $s$ in Eq.~(\ref{eq:aveDs}). As shown in Fig.~\ref{fig:averageDIs}(b), the behavior of $\mathcal{D}_s$ is almost similar for different $r$, such as an abrupt increase and then gradual decrease with $s$, resulting in the right-skewed unimodal curve. The decreasing pattern of DI with $s$ is expected and consistent with previous findings~\cite{giacomin2015}. On the other hand, $\mathcal{D}_s$ is also low for very short $s$ when the OD pairs are far from the city center. That is because in many routes of short distances, OD pairs are connected directly by single roads as shown in Fig.~\ref{fig:samplingOD_DI}(c). Many almost directly connected cases result in $\mathcal{D} \approx 1$~\cite{aldous2010}. Since scanning the travel routes in a radial manner to spatially dig up the DI, we can achieve this observation, namely the nonmonotonic behavior, unlike the previous findings. Meanwhile, the values of $\mathcal{D}_{s}$ at the fixed $s$ are different from each other depending on radius $r$. For example, $\mathcal{D}_{s}$ in $s$ = $20 \mathrm{km}$ ranges from the lowest around $1.3$ (when $r$ = $10 \mathrm{km}$) to the highest $1.7$ (when $r$ = $30 \mathrm{km}$). It describes the structural difference between the city center (smaller $r$) and the peripheral region (larger $r$). The $\mathcal{D}_s$ versus $s$ for the individual cities are investigated in Fig.~S2.

We also examine the DI in terms of the angular distance by measuring $\mathcal{D}_{\theta}$ [Eq.~(\ref{eq:aveDtheta})] as shown in Fig.~\ref{fig:averageDIs}(c). Note for $\mathcal{D}_s$ that the longer distance one travels, the less the route detours. Similarly, the decreasing pattern of $\mathcal{D}_{\theta}$ means that the route is less detoured when passing through the city center (large $\theta$) than when traveling along the peripheries (small $\theta$). It coincides with the previous finding that the route passing through the city center tends to be straighter and that the center effect exists universally in many cities~\cite{giacomin2015}. The peak is around $\pi/10$, which is about $18\sim20$ degree. Surprisingly, the $\mathcal{D}_\theta$ curves of different radii show a self-organized collapse on a single curve, except for the curve of the shortest radius ($r = 2 \mathrm{km}$) which has a slightly higher value. It means that regardless of the radius if routes have the same angular feature to the center, they are likely to detour with a similar ratio. The collapsed patterns of $\mathcal{D}_{\theta}$ are also found at the individual city level (see Supplemental Material for Fig. S4). Even though there are a little difference from city to city and more fluctuations than the averaged value over all cities, $\mathcal{D}_\theta$ curves of different radii are less fluctuate than the curves of $\mathcal{D}_s$ in the most cities.


The universal behavior of $\mathcal{D}_{\theta}$ guarantees the nice collapse of the $\mathcal{D}_{s}$ via the  trigonometric identity $s/r = 2 \sin( \theta/2) = \sqrt{2(1- \cos \theta)}$.  Thus, by introducing the scaled factor $r$ with the aid of this identity, one can express $\mathcal{D}_s$ with an arbitrary scaling function $f$ as
\begin{equation}
\mathcal{D}_{s}=f\left(\frac{s}{r}\right).
\label{eq:scaling}
\end{equation}
We plot $\mathcal{D}_{s}$ with respect to the scaled distance $s/r$ as shown in Fig.~\ref{fig:averageDIs}(d). 
It means that an urban road network has a seemingly different pattern of $\mathcal{D}_{s}$ for a given $r$, but $\mathcal{D}_{s}$ scales with the radial distance $r$ in a city, showing the inherently universal pattern. We also check the DI patterns from random samples where the route samples are not restricted to having the same radial boundary (see Supplemental Material for Fig.~S5 and Note 1). For the random samples, $\mathcal{D}_r\simeq 1.5$, and $\mathcal{D}_s$ follows the scaling form such as Eq.~(\ref{eq:scaling}). It signals that our observation of the universal pattern is not restricted to the sampling method.

All the aforementioned analyzes are repeated for the individual cities (see Supplemental Material for Table S1, Figs. S3 and S4, again). There is a city-wise fluctuation, but the entire tendencies for most individual cities also are similar.



\subsection{\label{sec:modelresult} Spatial core-periphery structure of road networks and universal DI pattern}

From empirical data analysis, we find the detourness in the road network has the radial invariant pattern, which are proven by the constant behavior of $\mathcal{D}_r$ and the clear collapse of $\mathcal{D}_{\theta}$ for various radii as illustrated in Figs.~\ref{fig:averageDIs}(a) and~\ref{fig:averageDIs}(c). Here, we further investigate how the radial invariant pattern is related to the spatial structure of the road network, especially focusing on the road density (the number of roads per area). The road density is known as a key factor determining detouring ratios. Empirical data shows that the DI and road density are negatively correlated (Fig. ~\ref{fig:statisticsDI}(d) and ~\cite{Levinson2009732, MERCHAN202038}), and we additionally confirm this relationship with our model network in the controlled condition (see Fig.~S1). 

The radial profile of the road density (precisely using the intersection density of the roads) describes that the road structure is denser near the central area, and the density decreases with the radius from the center (see Fig.S6(a)), which is similar to the other urban density gradient patterns in ~\cite{delloye2020alonso,lemoy2020evidence,anas1998urban,clark1951urban}. Such decreasing patterns are commonly found in almost all cities in our sample. Moreover, in the aggregated dataset over 70 cities (Fig.S6(a)) and many individual city data (Fig.S7), the decreasing patterns particularly follow the algebraic decay. In terms of functional aspect, we also find the central areas are better accessible to more diverse destinations through the network, than the peripheral areas (see Supplemental Material for Note 2 and Fig.~S6(b)). To sum up, the central area is better connected both physically and functionally, and the connectivity level decreases as going further from center. Such a radial structure shown in the road network seemingly evokes the notion of the core-periphery structure in network science~\cite{SHLee2014, SHLee2016}, where the core parts are densely connected with each other and peripheral parts are mostly connected to the cores (see the details in Supplemental Material for Note 2).

Taken together, we can expect that the detouring ratio of a route is the combined result of two intrinsic mechanisms, which are the spatial core-peripheriness of the network and the negative relationship between $s$ and DI. As the specific examples, routes in the central area would have the mixed effect of short distance (causing more detouring) and high road density (causing less detouring) on the DI. The two effects also coexist on routes located in the periphery but work in the opposite way as the long distance and low density. Eventually, the compensation of the two effects seems to make the detouring ratio at all radial positions similar to each other.

\begin{figure*}
\begin{center}
	\includegraphics[width=1.\linewidth]{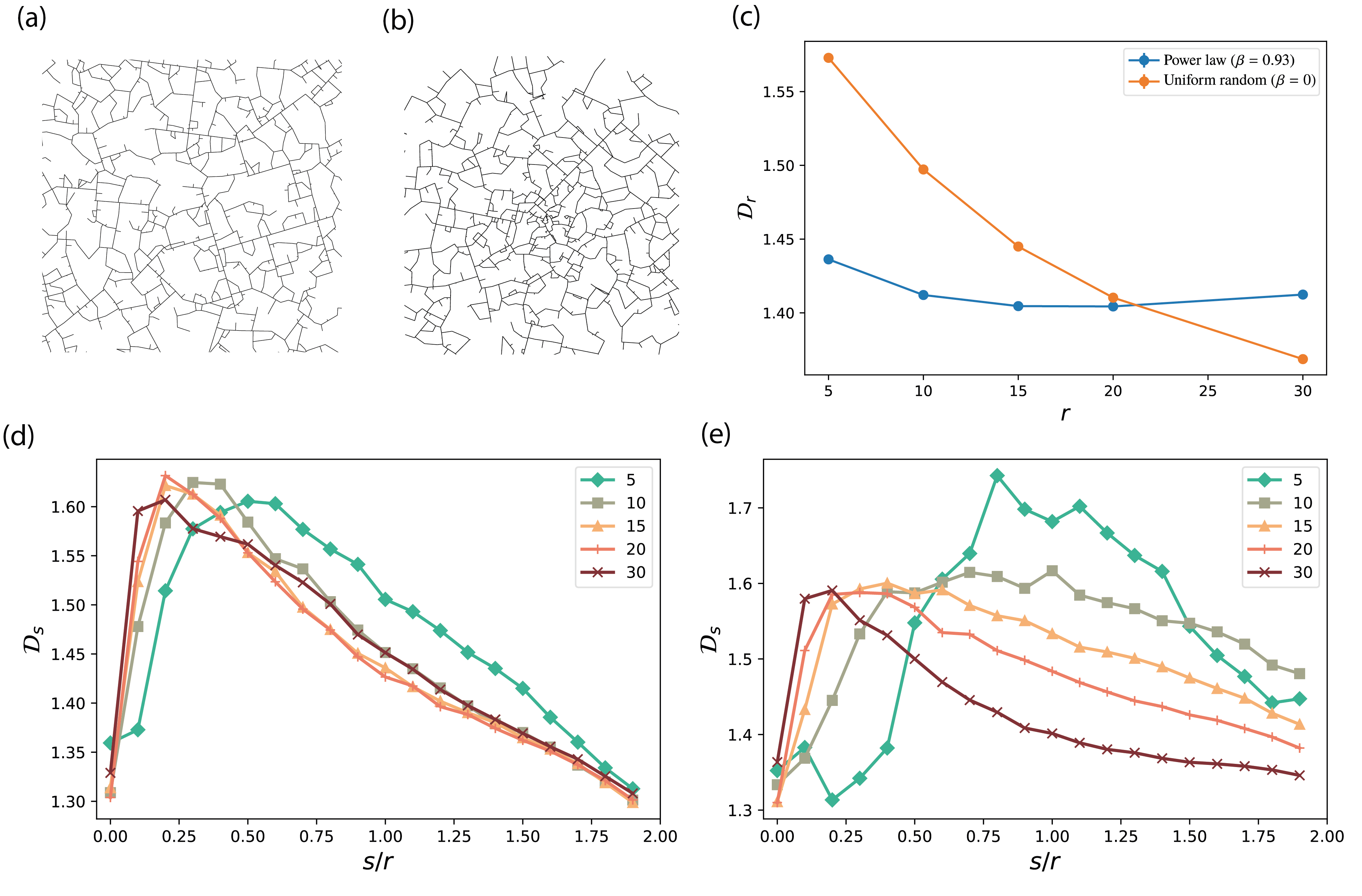}
\end{center}
\caption{Examples and the average detour indices of the artificial road networks. The examples are illustrated for (a) the uniform random distribution of the intersection ($\beta=0$) and (b) the power-law distribution of the intersection ($\beta=0.93$) where the $\beta$ is the exponent in $P(\vec{r})\sim r^{-\beta}$ [Eq.~(\ref{eq:position_powerlaw})]. The length threshold value is $h=4$. (c) The average value $\mathcal{D}_{r}$ is compared for the cases of the uniform distribution and the power-law distribution of the intersections. The average value $\mathcal{D}_{s}$ for radius $r=5$, 10, 15, 20, and 30 is plotted as a function of the scaled distance $s/r$ for (d) the power-law distribution and (e) the uniform distribution. One can detect that the artificial network with the power-law distribution of the intersection shows the constant behavior of $\mathcal{D}_r$ and the scaling collapse $\mathcal{D}_s$'s with the scaling factor $r$, whereas the network with the uniform random distribution does not. The results for $\mathcal{D}_{\theta}$ are shown in Fig.~S8.}
\label{fig:model_result}
\end{figure*}

We verify our assumption by intensive simulations on the artificial road networks. As described in Sec.~\ref{sec:modelnetwork}, we consider two types of network with the different spatial distribution $P(\vec{r})$ of the intersection: (i) the uniform distribution (corresponding to $\beta=0$) and (ii) the power-law distribution ($\beta>0$) [Eq.~(\ref{eq:position_powerlaw})]. In the case of the power law, we use the small values of $\beta$ (the average of $\beta$ is about $0.93$). Figures~\ref{fig:model_result}(a) and~\ref{fig:model_result}(b) illustrate the examples of the artificial road network generated by the distributions. At a glance, the network having the power-law distribution of intersections mimics the core-periphery-like structure. 

The primary constraints for the artificial road network are the planarity of the network and the cost-minimizing connection rule, and they are sufficient to reproduce the abruptly increasing and then decreasing pattern of $\mathcal{D}_s$ for both the model networks [Figs.~\ref{fig:model_result}(d) and~\ref{fig:model_result}(e)]. However, the constant behavior of $\mathcal{D}_r$ only emerges in the network using the power-law distribution, i.e., the core-peripheral structure [Fig.~\ref{fig:model_result}(b)]. Moreover, the constant value $\mathcal{D}_r\simeq 1.42$ is almost the same as the empirical value $\mathcal{D}_r\simeq 1.5$. The $\mathcal{D}_r$ in the randomly distributed network shows the decreasing behavior. Especially, the $\mathcal{D}_r$ at small $r$ is larger than that in the power-law case. Due to an absence of the densely connected core in the randomly distributed network [Fig.~\ref{fig:model_result}(a)], the network does not have enough shortcuts to lower the DI of the OD pair with a short distance in central areas. In contrast, for the distant OD pair in the periphery, the effect of reduced road density is missing in the random network, thus DI keeps lowering under the effect of long distance solely.

As mentioned above, the $\mathcal{D}_s$ in both model networks shows the qualitatively same behavior (the increasing and then decreasing pattern) as the empirical data, but they have a clear difference in the viewpoint of the scaling. In Fig.~\ref{fig:model_result}(d), the scaling curve of $\mathcal{D}_s$ in the network with the power-law distribution collapses to a single curve, and a form of a scaling function may seem similar to $f$ in Eq.~(\ref{eq:scaling}). The position of the peak is about $s/r\simeq0.25$ which is also observed in the empirical data [Fig.~\ref{fig:averageDIs}(d)]. On the contrary, the $\mathcal{D}_s$ with $s/r$ in the randomly distributed network does not collapse into a single curve but is rather scattered [Fig.~\ref{fig:model_result}(e)]. The core-periphery structure of the road network has not been considered in the previous studies for the road network analysis, but our finding reveals that the spatial structure is contributed to the universal scaling behavior of $\mathcal{D}_s$ including the constant $\mathcal{D}_r$.

\subsection{\label{sec:application} City-specific detour patterns}
\begin{figure}
\begin{center}
\includegraphics[width=1\linewidth]{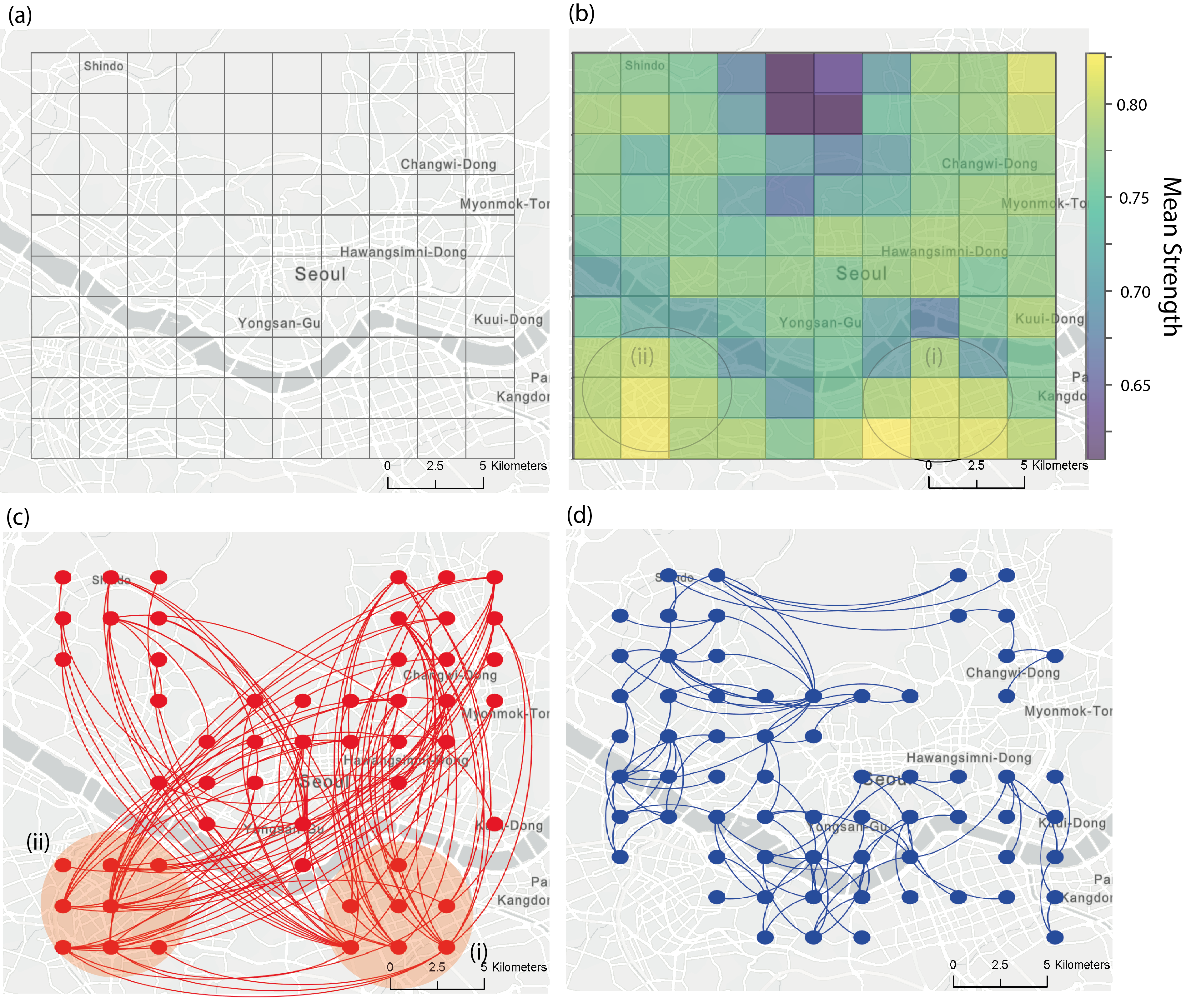}
\end{center}
\caption{The detour index network of Seoul. (a) The map of Seoul is divided into $10 \times 10$ regions. A centroid of each region or each cell corresponds to a node. (b) The color encodes the region efficiency $R_i$ of the region $i$ [Eq.~(\ref{eq:region_efficiency})] where $R_i$ is the sum of the efficiency level of edges $E_{ij}$ defined as Eq.~(\ref{eq:edge_efficiency}). The brighter the color, the more efficient the area is. (c) The 100 most efficient edges and (d) the 100 least efficient edges are shown, and they are selected by the 100 highest and 100 lowest $E_{ij}$'s, respectively. The two efficient areas, (i) Gangnam and (ii) Yeongdeungpo are highlighted by either empty circles with black solid lines in panel (b) or red-shaded circles in panel (c).} 
\label{fig:local_efficiency_map}
\end{figure}

Finally, we visualize the spatial pattern of the DI for the road networks of a city to take a deeper look at the city-specific structure. We analyze the 20 $\mathrm{km}$ $\times$ 20 $\mathrm{km}$ square size of Seoul as a representative case study in the main text and also include the spatial maps for several cities (see Supplemental Material for Figs. S9-S14). The course-grained DI network in Fig.~\ref{fig:local_efficiency_map} is proposed to directly show which regions are efficiently connected to the other parts of the city. For visualization, we divide the city map into 10 by 10 regions (cells) as shown in Fig. \ref{fig:local_efficiency_map}(a) and consider an all-to-all network whose node is the centroid of each region (each cell) and an edge weight is an efficiency level between an associated pair. The efficiency level assigned to an edge between the regions $i$ and $j$ is defined as
\begin{equation}
E_{ij}\equiv \left(\frac{1}{N_{\rm OD}^{\prime}}\sum_{\rm O, D}^{\prime} D_{\rm OD}\right)^{-1}=\left(\frac{1}{N_{\rm OD}^{\prime}}  \sum_{\rm O, D}^{\prime} \frac{d_{\rm OD}}{s_{\rm OD}}\right)^{-1},
\label{eq:edge_efficiency}
\end{equation}
where the primed summation is for the pair (O, D), with O (D) belonging to either region $i$ $(j)$ or $j$ $(i)$, and $N_{\rm OD}^{\prime}$ is the number of the pairs. According to the formula in Eq.~(\ref{eq:edge_efficiency}), the large $E$ is read as the high-efficient connectivity. To compute the efficiency, we randomly sample the $50\,000$ OD pairs (recall that the universal pattern of the DI in our analysis is independent of the sampling method). The region efficiency for a region $i$ is also defined as
\begin{equation}
    R_{i} \equiv \frac{1}{k_i}\sum_{{\textrm{n.n. of }}i} E_{ij},
\label{eq:region_efficiency}
\end{equation}
namely, the mean node-strength [the average of the edge weights $E$ in Eq.~(\ref{eq:edge_efficiency}) over $k_i$ nearest neighbors of the node $i$ (almost $k_i\simeq 10 \times 10$) here] and is illustrated in Fig.~\ref{fig:local_efficiency_map}(b). The higher $R$ is colored brighter, which indicates that the node is connected to other places via efficient routes. The visualization in Fig.~\ref{fig:local_efficiency_map}(b) captures such a behavior well throughout Seoul. As the most efficient areas, the two brightest areas (i) Gangnam and (ii) Yeongdeungpo are shown on the bottom enclosed by large circles. They are located at the corners, which allows them to have longer Euclidean distance pairs and a lower DI, so it seems natural for them to be considered efficient areas.

However, we can also capture the local characteristics revealed by the spatial DI pattern in Fig.~\ref{fig:local_efficiency_map}(b). If the Seoul road structure were exactly isotropic, the four corners of the boundary would be equally the most efficient areas, but it does not in our observation. The disparity in node strengths between the four corners may stem from the intrinsic socioeconomic and geographic characteristics of Seoul. The Gangnam area is known as a new subcenter of Seoul, receiving great benefit from infrastructure investment, and the Yeongdeungpo area is also likely to enjoy the benefit of good transportation infrastructure, as it is located next to the Yeoyido that is another subcenter of Seoul~\cite{KIM2012142}. The two northern corners are not prominent compared to Gangnam and Yeongdeungpo. The areas are not as economically active as the southern areas, but rather play the role of residential areas. In addition, the darkest areas indicating low $R_i$ in the middle of the city and across the city are natural barriers, Bukhansan (mountain) and  Hangang (river), respectively. The region efficiency $R_i$ can provide a hint of the role of the region, such as a subcenter or a major transportation hub. 

Note that the efficiency level can be comparable to the closeness centrality~\cite{NewmanBook}, defined mainly as the reciprocal of the total shortest distances (strictly the shortest hopping distance), i.e., $\left(\sum d_{\rm OD}\right)^{-1}$. In terms of closeness centrality, the region on the corner such as Gangnam and Yeongdeungpo can have perhaps a smaller value than the center. It may mean that the region with the smaller closeness has low proximity and then low efficiency of the travel routes. One recognizes that the closeness centrality and the DI capture efficiency of travel routes from different viewpoints (considering the shortest path only for the former and the ratio of the shortest and Euclidean distance for the latter), leading to the different interpretations of a travel efficiency in a different context. 

Now we select the 100 most efficient pairs (highest $E_{ij}$'s) and the 100 least efficient pairs (lowest $E_{ij}$'s) and display in Figs.~\ref{fig:local_efficiency_map}(c) and \ref{fig:local_efficiency_map}(d), respectively, to identify the spatial pattern of efficient connections. The highly efficient edges usually traverse the center, which accords with the previous observation of the average pattern. Interestingly, Gangnam [highlighted by a red circle with label (i) in Fig.~\ref{fig:local_efficiency_map}(c)] is efficiently connected to areas in various directions, such as left-top corners and even right-top areas. It reveals that Gangnam is a particularly efficient area (already demonstrated by the high $R_i$) that is accessible from other areas of Seoul. It is understandable according to the history of land development in Seoul. The Seoul government has focused on investing in road infrastructure in the Gangnam area through various city policies, such as the land readjustment project to disperse the population overcrowded in the northern part of Seoul and to build the area as a new sub-center since the 1980's~\cite{KIM2012142}. This government involvement has improved accessibility to and within the Gangnam area and generated a special topological characteristic in the Seoul road network. The 100 least efficient edges are illustrated in Fig.~\ref{fig:local_efficiency_map}(d). We omit the edges whose nodes reached following the edges are located on Bukhansan on purpose because they have trivially low values of $E_{ij}$ due to the impact of the natural barrier. The selected low-efficient connections are mostly short-distance OD pairs as expected, giving the generally short routes, and spread over the whole area.

\begin{figure}
\begin{center}
\includegraphics[width=1\linewidth]{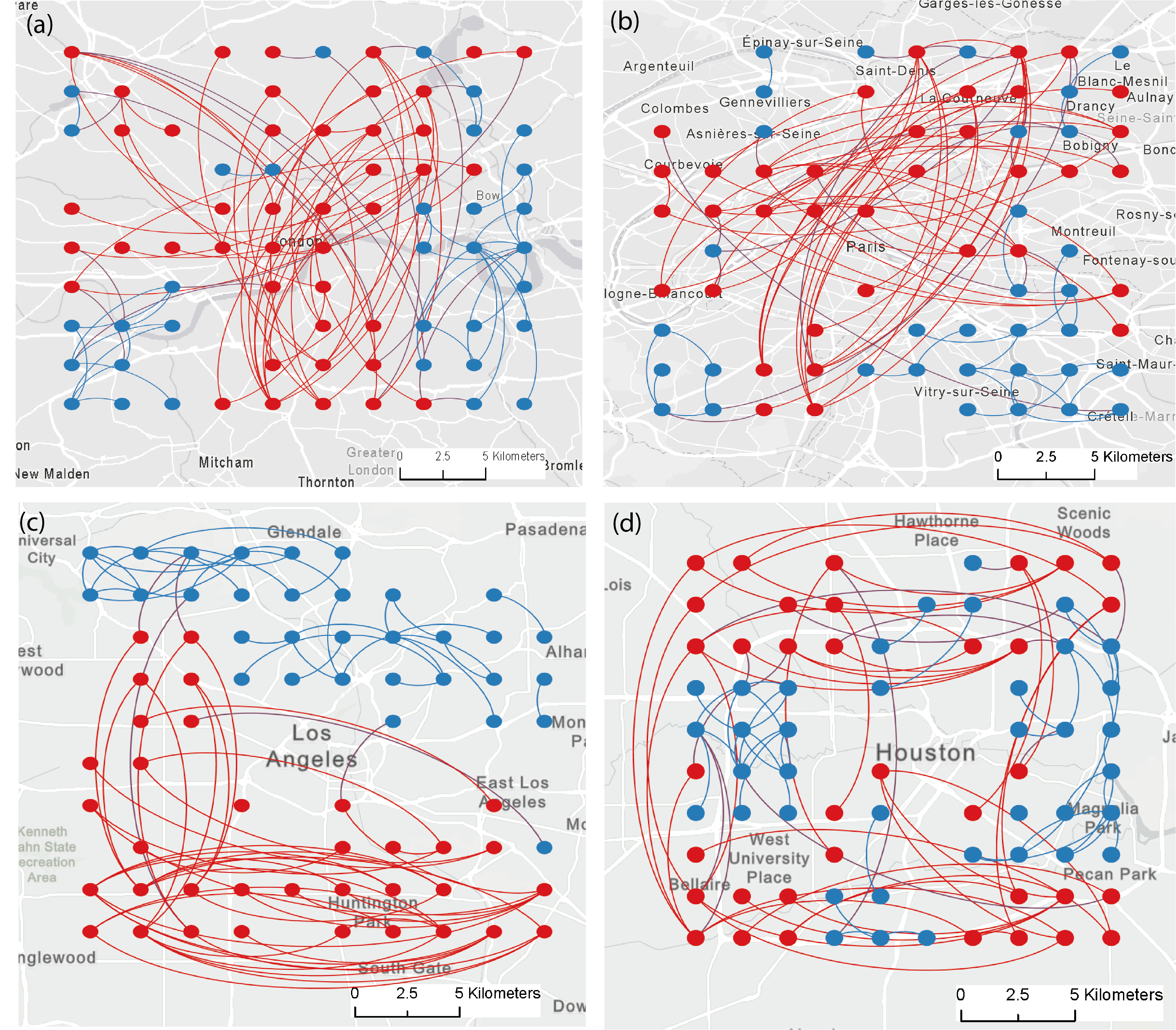}
\end{center}
\caption{The detour index network on the city map of (a) London, (b) Paris, (c) Los Angeles, and (d) Houston. The networks only show the 50 most efficient edges (red lines) and the 50 least efficient edges (blue lines) and the relevant nodes.} 
\label{fig:local_efficiency_map_4}
\end{figure}

Finally, we show the additional DI networks with the 50 high-efficiency (red) and low-efficiency (blue) edges in the four global cities such as London, Paris, Los Angeles, and Houston in Figs.~\ref{fig:local_efficiency_map_4}(a)-\ref{fig:local_efficiency_map_4}(d). London and Paris are representative metropolitan areas in Europe that have evolved from the center for many years with a long history, despite their current polycentric properties~\cite{doi:10.1080/02723638.2014.939538}. Houston and Los Angeles are also representative American cities, known as planned cities without distinct centers~\cite{doi:10.1080/02723638.2016.1200279, GLAESER20042481}. In the cities like London, Paris, and Seoul, the efficient connections pass through the centers. This implies that road networks have been developed to connect central areas with other parts of the city, similar to the spirit of the core-peripheriness. However, American cities do not show such patterns. In Los Angeles, the efficient connections are spread out only in half region of the sampled area without any special pattern [Fig.~\ref{fig:local_efficiency_map_4}(c)]. In Houston, the 50 most efficient pairs lie along the highway in the outer area [Fig.~\ref{fig:local_efficiency_map_4}(d)].  The patterns indicate that the geographical center in the city has not played a pivotal role to provide an efficient travel route in the road structure, but the peripheral areas are more equally connected with each other. The course-grained DI network is simple to represent the road structure, but it allows us to abstract and visualize the local characteristic of urban street networks. One can note that the five cities of Seoul, London, Paris, Los Angeles, and Houston have very similar values of the $\langle D \rangle_{\rm fix}$ and $\langle D \rangle_{\rm ran}$ between $1.2$ and $1.4$ in Fig.~\ref{fig:statisticsDI}(b) as well as the similar collapse behavior of $\mathcal{D}_{\theta}$ as seen in Fig.~S4. The macroscopic phenomena do not inform us of the city-specific structures as provided in Figs.~\ref{fig:local_efficiency_map} and~\ref{fig:local_efficiency_map_4}. This analysis of the DI networks in parallel with the macroscopic observation is definitely useful to understand the road structures for individual cities.

\section{\label{sec:discussion}Summary and Discussion}
Numerous studies on urban road networks have focused on macroscopic phenomena such as scaling behavior between relevant quantities. So far, less attention has been paid to the internal spatial pattern of the networks, so we have studied the spatial pattern using the detour index (DI). The DI representing the travel-route efficiency is the practical measure in the analysis of the road network that indicates how straightly one can access the destination contingent upon the given road network and thus reflects the perspective of the traveler on the structure of the road network. Consequently, the spatial pattern of the DI signifies the efficiency of the travel route based on the user experience in different parts of urban areas. 

In this study, we have empirically analyzed the behavior of the DI upon the spatial positions of the routes for the 70 global cities by introducing the radial approach (leading to the radius-fixed sampling). The investigation into the spatiality of the DI toegether with the radial approach has enabled us to find the relationship between urban spatial property and the efficiency of road networks. The power-law distribution of the intersection's position from the center is related to the universal behavior of the DI: the constant behavior of $\mathcal{D}_r$  [Fig.~\ref{fig:averageDIs}(a)] and the collapse into a single curve of $\mathcal{D}_{\theta}$ [Fig.~\ref{fig:averageDIs}(c)]. The nice collapse of $\mathcal{D}_{\theta}$ provides the scaling factor $r$ for $\mathcal{D}_s$, i.e., $\mathcal{D}_s=f(\frac{s}{r})$ [Fig.~\ref{fig:averageDIs}(d)]. With the help of the artificial model networks, we have verified that the spatial pattern of the intersection density having the core-periphery structure (i.e., characterized by a power-law distribution, rather than a uniform distribution) is likely to be responsible for the universal behavior of the DI [see Fig.~\ref{fig:model_result}]. Considering that previous studies have rarely dealt with the spatial structure of road networks, our finding is quite intriguing in the sense that it provides a clue to the connection between this structure and the macroscopic behavior.  

Although we have discovered the relation between the averaged DI as a macroscopic quantity and the intrinsic spatial pattern characterized by the intersection distribution, the average value has a limitation to reveal the city-specific structure. To look into the intra-city structure, we have proposed the visualization method of the DI networks by introducing the efficiency level as an edge weight, yielding the region efficiency as a mean node-strength. Basically, the connectivity pattern usually depends on the characteristics of the city, such as the (self-organized) historical or planned city [Figs.~\ref{fig:local_efficiency_map} and~\ref{fig:local_efficiency_map_4}]. Historical cities like Seoul, London, and Paris clearly show the strong effect of the centers demonstrated by the fact that the most efficient connections pass through the center, but planned cities like US cities do not. The difference is also consistent with our understanding of the urban structure of these cities. For example, in planned cities, such as US cities, the population density gradient is much less pronounced than in Asian and European cities~\cite{rodrigue2020geography}.

Our analysis can be extended to diverse considerations, such as an analytic approach to the spatial distribution of intersection and the universal pattern of the DI, the position of the center and the polycentric structure, and the leverage of the core-periphery structure in the DI pattern, which will be explored in future work. Nonetheless, this study becomes a basic step in broadening an understanding of the effect of spatial characteristics (e.g., the distribution of the intersection density) on the universal pattern in the urban road networks (e.g., the scaling behavior of the DI). We also expect that consideration of spatial position in road network analysis would bring more practical implications, which are useful to our real life, as well as deepen theoretical understanding.

\section*{Acknowledgments}

This research was supported by the Korea National Research Foundation (NRF) through grant numbers. NRF-2020R1A2C2010875 (S.-W.S.) and NRF-2021R1C1C1007918 (M.J.L.). This work was also supported by "System Dynamics Simulation Approach to Competitive Dynamics, Environments and Employment Impacts of Small Businesses in Urban Commercial Areas" through the Ministry of Education of the Republic of Korea and NRF grants funded by the Korea government (MSIT) (No.2020R1A2C2008443) (SH.C.). This work was also partly supported by Institute of Information \& communications Technology Planning \& Evaluation (IITP) grant funded by the Korea government(MSIT) (No.RS-2022-00155885, Artificial Intelligence Convergence Innovation Human Resources Development (Hanyang University ERICA)) (S.-W.S.). We also acknowledge the hospitality at APCTP where part of this work was done.

%


\end{document}